\documentclass{ws-procs9x6}

\def\1s0{{}^1S_0}
\def\ts1{{}^3S_1}
\def\td1{{}^3D_1}
\newcommand{\Slash}[1]{\ooalign{\hfil/\hfil\crcr$#1$}}
\def\piless{EFT($\Slash{\pi}$)}
\def\pieft{EFT($\pi$)}

\begin{document}

\title{Renormalization group analysis\\
of nuclear current operators}

\author{Satoshi X. Nakamura}

\address{Theory Group, TRIUMF,
4004 Wesbrook Mall, Vancouver, BC V6T 2A3, Canada\\
E-mail: snakamura@triumf.ca}

\author{Shung-ichi Ando}

\address{Department of Physics, Sungkyunkwan University,
Suwon, 440-746 Korea\\
E-mail: sando@meson.skku.ac.kr}

\begin{abstract}
We review our study of  
the Wilsonian renormalization group (WRG) analysis 
for nuclear current operators. % in two-nucleon system.
We apply WRG method to axial-current operators
derived from various approaches
and obtain 
%by integrating out high-energy modes,
%by reducing the model space of the operators,
the unique effective
low-energy operator.%, $O_{low\mbox{\rm -}k}$.
\end{abstract}


\bodymatter

\vspace{7mm}
\noindent
In Wilsonian renormalization group (WRG) analysis,
one integrates out high-energy modes and examines the evolution of
interactions.
We apply the WRG analysis to nuclear operators such as nuclear
potentials and current operators.
It is known that various nuclear potentials 
equally well reproduce all of the data
below the pion production threshold,
while they
appear quite different in describing 
the short range part. 
As a result of the model-space reduction using WRG equation, all
the potentials converge to a single effective low-momentum
potential.~\cite{ref1}.  %, {\it i.e.}, $V_{low\mbox{-}k}$
Moreover, a parameterization of the single 
potential becomes the NEFT-based operator
which, by construction, does not 
depend on modeling the small scale physics.

In evaluating an % transition 
amplitude of an electroweak
process in few-nucleon system,
nuclear %force and 
current operators are necessary ingredients 
as well as the nuclear force.
The %nuclear 
current 
operators based on different approaches
have quite different behaviors in the short-range part,
however, all of them give essentially the same reaction rates
for low-energy reactions,
{\it e.g.}, solar-neutrino reactions on the deuteron~\cite{ref2}.
This implies that we can obtain a single effective
current operator %, {\it i.e.} $O_{low\mbox{-}k}$, 
through the WRG analysis. 

\begin{figure}[t]
\begin{center}
\psfig{file=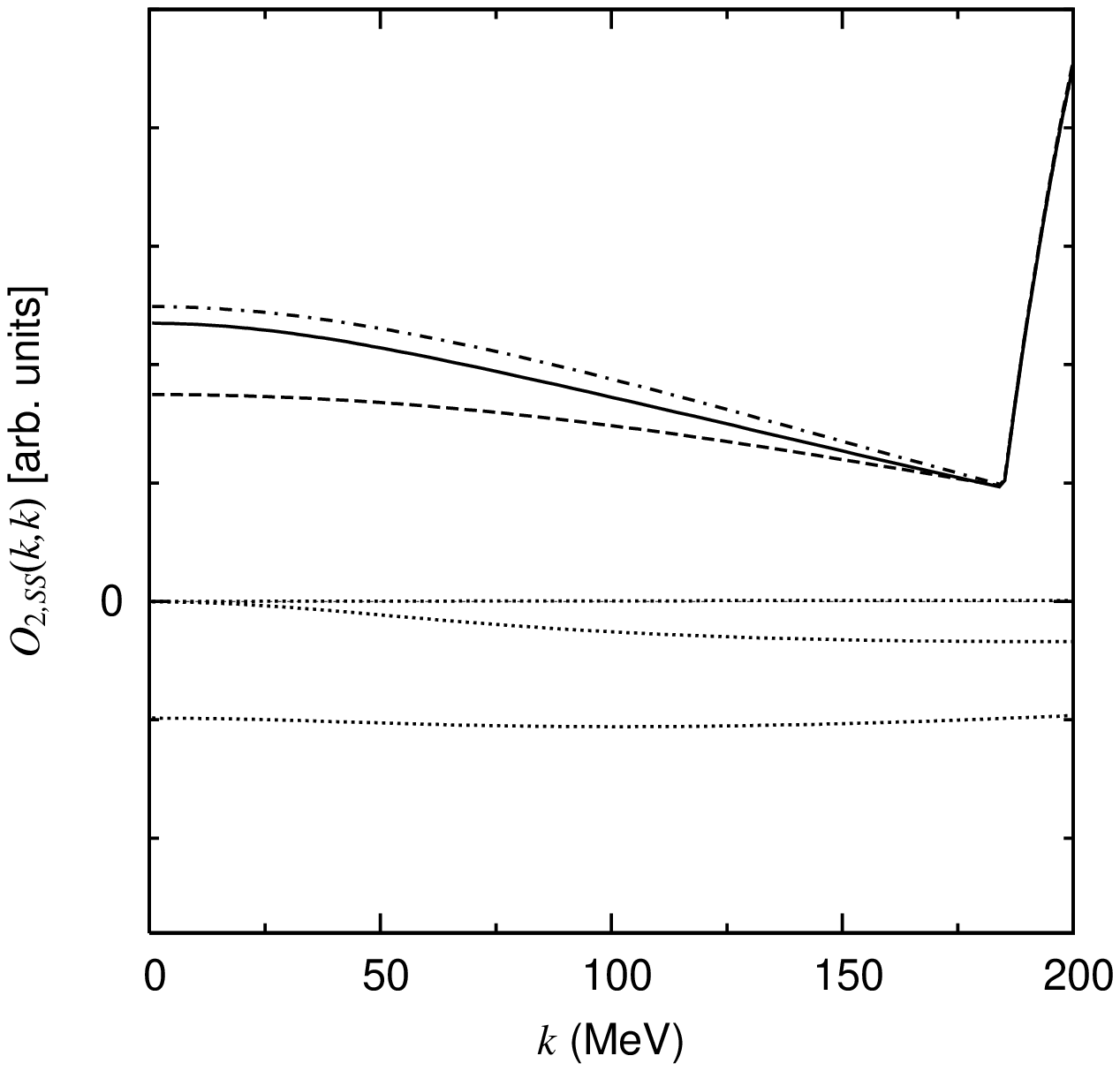,width=5.6cm}
\psfig{file=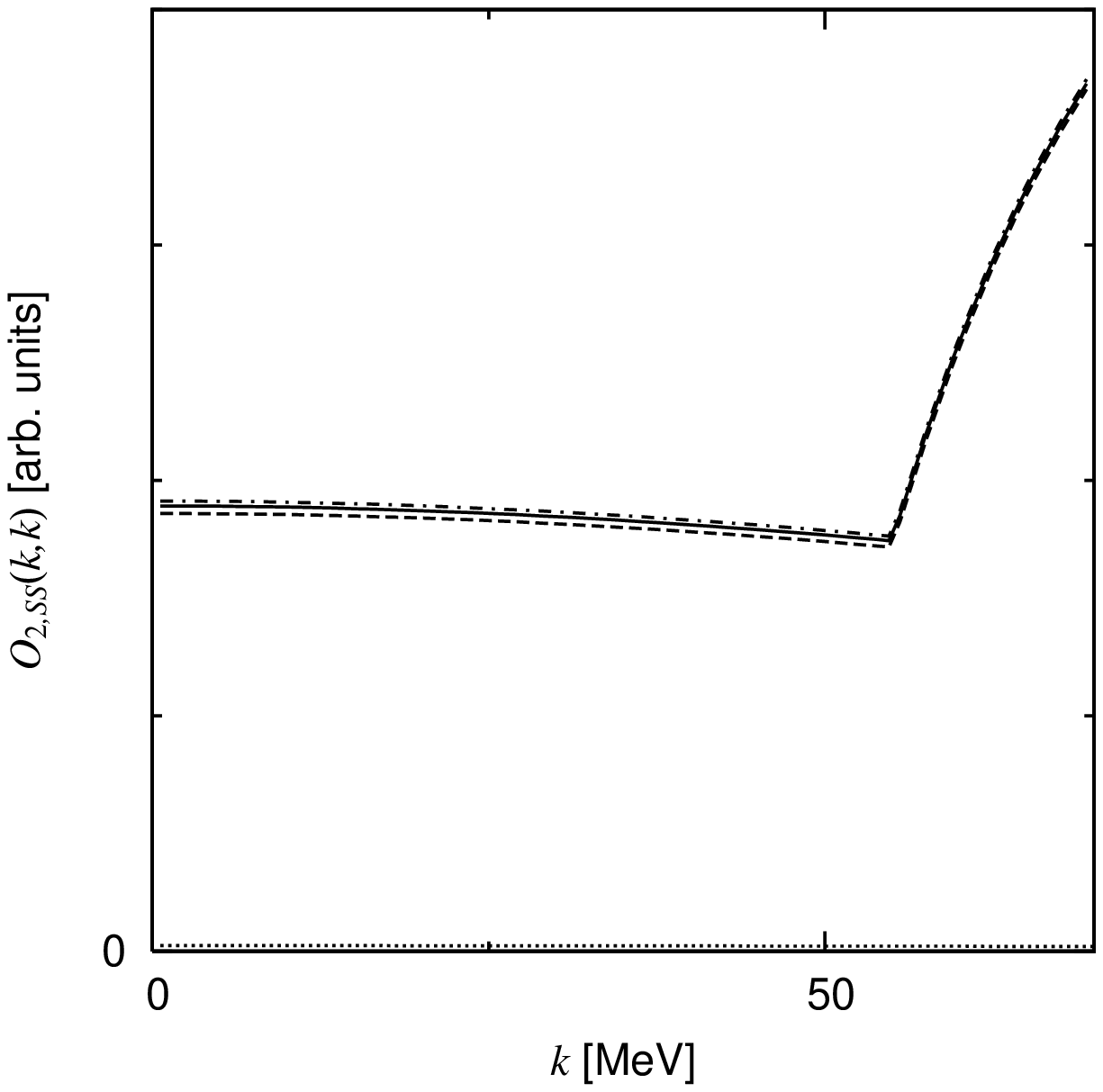,width=5.6cm}
\end{center}
\caption{
Effective two-body current operators for initial $S$- to final $S$-wave 
state evaluated with the cutoff $\Lambda = 200$ MeV (left panel) 
and 70 MeV (right panel).
}
\label{fig;olowk}
\end{figure}

We derive a WRG equation for the current operator 
and use it %
to reduce the model space of currents
from the models or NEFT with the pion (\pieft)~\cite{ref3}.
We are specifically concerned with the exchange axial-current
operators relevant to the solar neutrino-induced breakup 
of the deuteron.

In Fig.~\ref{fig;olowk}, 
by reducing the model space, %of the operators 
we find evolution of the bare two-body operators (lower three curves)  
to the effective ones (upper three ones). 
However, a model dependence still remains
among the effective operators %from the models and \pieft\ at
at $\Lambda$ = 200 MeV (left panel).
This is because even the one-pion range mechanism is model dependent.
These effective currents are further evolved up to 
$\Lambda=70$ MeV (right panel).
With this resolution of the system, 
the model dependence among the %bare 
currents
is not seen any more, and thus we
obtain the unique effective operator. % $O_{low\mbox{\rm -}k}$.

Furthermore, we simulate the effective two-body current with
$\Lambda=70$ MeV using the \piless-based
parameterization.
Except for ``jump-up'' structures in Fig.~\ref{fig;olowk} 
due to the bare one-body current contribution,
the two-parameter fit yields an almost perfect simulation.
Therefore, one can obtain
the \piless-based %current 
operator %can be obtained 
from the models or \pieft\ in this way.

%\footnote{
%SXN is supported by 
%the Natural Science and Engineering Research Council of Canada. 
SA is supported by  
Korean Research Foundation and The Korean Federation of Science
and Technology Societies Grant funded by Korean Government 
(MOEHRD, Basic Research Promotion Fund). 
%}

\end{document}